\documentclass[final]{aipproc}

\layoutstyle{6x9}

\usepackage{mathrsfs}

\usepackage{dsfont}

\usepackage{amsmath, amsthm, amssymb}

\begin{document}

\title{Semiclassical Analysis of Constrained Quantum Systems}

\classification{03.65.Sq, 03.65.Pm, 98.80.Qc}

\keywords {Dirac's constraint quantization, semiclassical states,
effective equations}

\author{Artur Tsobanjan}{
address={Institute for Gravitation and the Cosmos, The Pennsylvania
State University, 104 Davey Lab, University park, PA 16802 \\
E-mail: axt236@psu.edu}}

\begin{abstract}
Exact procedures that follow Dirac's constraint quantization of
gauge theories are usually technically involved and often difficult
to implement in practice. We overview an ``effective'' scheme for
obtaining the leading order semiclassical corrections to the
dynamics of constrained quantum systems developed elsewhere.
Motivated by the geometrical view of quantum mechanics, our method
mimics the classical Dirac-Bergmann algorithm and avoids direct
reference to a particular representation of the physical Hilbert
space. We illustrate the procedure through the example of a
relativistic particle in Minkowski spacetime.
\end{abstract}

\maketitle

\section{Introduction}

Full phase-space of fundamental gauge theories does not represent
the physical degrees of freedom: constraint functions restrict the
part of the phase-space that is physically accessible and the
transformations they generate identify points that are physically
equivalent. Direct quantization of the physical degrees of freedom
requires global knowledge of the solutions to the constraint
functions and the corresponding gauge orbits which, in general is
unavailable. In its absence one can follow a method due to
Dirac~\cite{DirQuant}, which postpones enforcement of the
constraints until after quantization. Constraints are to be promoted
to quantum operators and solved by demanding that they annihilate
physical states. Often the most challenging part of this procedure
is determining the physical inner product on the space of solutions.
Explicit constructions of the physical Hilbert space are only
available in few specific cases.

Here we overview an ``effective'' scheme for solving constraints at
the quantum level developed in~\cite{EffCons} and~\cite{EffConsRel}.
Focusing on kinematically semiclassical states we obtain
quantitative description of the leading order quantum corrections
without a rigorous definition of the physical inner product. In its
current form, the method applies to theories with finite number of
classical degrees of freedom, such as symmetry-reduced
minisuperspace models. In each section of the report we illustrate
the corresponding part of our construction using the example of a
free relativistic particle.

\section{Kinematical Representation}

According to Dirac's prescription, one is first to quantize the free
system to obtain what is commonly referred to as a \emph{kinematical
representation}. We assume that in this process some closed Poisson
algebra with a finite number of generators, that completely
describes the classical phase-space, is quantized canonically.
Explicitly, there are N real phase-space functions $a_i$, $i=1,
\ldots {\rm N}$\ that are closed with respect to the Poisson bracket
$\{a_i, a_j\} = \sum_k \alpha_{ij}^{\ \ k} a_k$, where
$\alpha_{ij}^{\ \ k}$\ are structure constants. Quantization
identifies these with self-adjoint operators $\mathbf{a}_i$, whose
commutation relations represent the quantized version of the
classical Poisson bracket $[ \mathbf{a}_i, \mathbf{a}_j ] :=
\mathbf{a}_i \mathbf{a}_j - \mathbf{a}_j \mathbf{a}_i = i \hbar
\sum_k \alpha_{ij}^{\ \ k} \mathbf{a}_k$. For our purposes we assume
that the kinematical operators generate an associative unital
algebra $\mathscr{A}$\ containing all finite polynomials in
$\mathbf{a}_i$. The issue of domains of definition of unbounded
operators may be kinematically addressed by using a
non-Cauchy-complete pre-Hilbert inner product space representation.
This algebra has a countable linear basis $\mathbf{a}_{\rm
1}^{n_{\rm 1}}\mathbf{a}_{\rm 2}^{n_{\rm 2}} \ldots \mathbf{a}_{\rm
N}^{n_{\rm N}}$---due to the commutation relations, other orderings
of these polynomials are related through addition of lower order
polynomial terms.

We further assume that there is a single constraint $\mathbf{C}$,
which is a given element of $\mathscr{A}$. The case of several
commuting constraints may be treated in an analogous manner. In
order to sidestep the issue of the physical inner product we will
avoid direct reference to a Hilbert space, instead a \emph{state} in
our construction is a complex-valued linear functional on
$\mathscr{A}$. This is a more general notion than a Hilbert-space
state, in particular it includes density states and states that are
non-positive with respect to the kinematical $\star$-relations. A
state can be specified completely by specifying the values it
assigns to a linear basis on $\mathscr{A}$. We restrict to states
that assign $1$ to the identity element---a rather weak version of
normalization.

\emph{Free Relativistic Particle.} The classical phase-space of a
relativistic point particle of rest-mass $m$\ in Minkowski spacetime
with one space and one time dimension is coordinatized by two
canonical pairs $t, p_t$\ and $q, p$ with the non-zero Poisson
brackets $\{t,p_t\} = 1 = \{q, p\}$. The dynamics is governed by a
single constraint
\[
C = p_t^2 - p^2 - m^2\,.
\]
The standard quantum kinematical representation of this system is
given by the action of differential operators on smooth
wavefunctions in two variables $x_0$\ and $x_1$:
\[
\mathbf{t} = x_0 \quad , \quad \mathbf{p}_t = \frac{\hbar}{i}
\frac{\partial}{\partial x_0} \quad , \quad \mathbf{q} = x_1 \quad ,
\quad \mathbf{p} = \frac{\hbar}{i} \frac{\partial}{\partial x_1}\,.
\]
The associative algebra these generate has a linear basis
$\mathbf{t}^k \mathbf{p}_t^l \mathbf{q}^m \mathbf{p}^n$---terms with
a different ordering may be expressed using the well-known
commutation relations
\[
[\mathbf{t}, \mathbf{p}_t] = i\hbar \mathds{1} \quad, \quad
[\mathbf{q}, \mathbf{p}] = i \hbar \mathds{1}\,.
\]
Henceforth we will no longer refer to a concrete representation of
$\mathscr{A}$\ in terms of operators on a vector space---unless
stated otherwise, we will treat a state as an assignment of a
complex number to each basis polynomial. There is no
product-ordering ambiguity in the case of the above constraint and
quantization identifies it with the element $\mathbf{C} =
\mathbf{p}_t^2 - \mathbf{p}^2 - m^2 \mathds{1}$\ of $\mathscr{A}$.

\section{Quantum Variables}

In the previous section we have defined a state as a complex linear
functional on $\mathscr{A}$, which is then an element of its
vector-space dual, henceforth referred to as $\bar{\mathscr{A}}$. We
will denote the value assigned by the state $\psi \in
\bar{\mathscr{A}}$\ to the element $\mathbf{A} \in \mathscr{A}$\ by
$\langle \mathbf{A} \rangle_{\psi}$. Dropping the reference to a
particular state we write $\langle \mathbf{A} \rangle$\ to denote
the corresponding complex-valued function on the space of all
states. In particular, the set of functions induced by the linear
basis $\langle \mathbf{a}_{\rm 1}^{n_{\rm 1}}\mathbf{a}_{\rm
2}^{n_{\rm 2}} \ldots \mathbf{a}_{\rm N}^{n_{\rm N}} \rangle$
completely coordinatizes the space $\bar{\mathscr{A}}$.

Functions on the space of states inherit a Poisson bracket from the
commutation relations on $\mathscr{A}$
\[
\{ \langle \mathbf{A} \rangle, \langle \mathbf{B} \rangle \} :=
\frac{1}{i \hbar} \left\langle [ \mathbf{A}, \mathbf{B} ]
\right\rangle\,.
\]
The bracket satisfies the Jacobi identity and can be extended to
non-linear functions by using Leibnitz's rule. A flow induced on
observables by a Hamiltonian operator may be expressed entirely
using this bracket. In ordinary quantum mechanics the evolution of
the expectation value of an observable is governed by the equation
\[
\frac{\mathrm{d}}{\mathrm{d}t} \langle \mathbf{O} \rangle =
\frac{1}{i\hbar} \left\langle \left[ \mathbf{O}, \mathbf{H} \right]
\right\rangle +\frac{\partial\langle\mathbf{O}\rangle}{\partial
t}\,.
\]
By our earlier definition, the first term on the right-hand side is
simply the quantum Poisson bracket $\{\langle \mathbf{O} \rangle,
\langle \mathbf{H} \rangle \}$. The evolution of any function on
$\bar{\mathscr{A}}$\ induced by the Hamiltonian can be expressed
using a generalization of this equation
\[
\frac{\mathrm{d}}{\mathrm{d}t} f = \{ f, \langle \mathbf{H} \rangle
\} +\frac{\partial f}{\partial t}\,.
\]
This construction is related to the various geometrical formulations
of quantum mechanics~\cite{Schilling}~\cite{Garbowski_Kus_Marmo}\
and can be used to replace the single partial-differential
Schr\"odinger equation by an infinite number of coupled ordinary
differential equations, which can be of advantage in certain regimes
\cite{EffAc}.

To investigate semiclassical properties, we use coordinates on
$\bar{\mathscr{A}}$, that are adapted to the semiclassical
expansion, namely the $N$\ expectation values $\langle \mathbf{a}_i
\rangle$\ and the infinite number of moments $\left\langle
(\mathbf{a}_{\rm 1}-\langle \mathbf{a}_{\rm 1} \rangle)^{n_1} \ldots
(\mathbf{a}_{\rm N}-\langle \mathbf{a}_{\rm N} \rangle)^{n_N}
\right\rangle_{\rm Weyl}$, where the subscript ``Weyl'' indicates
the total symmetrization of terms in the product. Semiclassical
states are close to maximum localization in the phase-space
variables (bounded from below by the various uncertainty relations).
We assume that for such a state the moments fall off as
$\hbar^{\frac{1}{2}\sum_{i=1}^N n_i}$, this assumption can be
concretely realized using Gaussian wavefunctions.

\emph{Free Relativistic Particle.} The system's quantum state is
completely determined by the values of $\langle \mathbf{t}^k
\mathbf{p}_t^l \mathbf{q}^m \mathbf{p}^n \rangle $, or alternatively
the expectation values of the four basic observables $\langle
\mathbf{t} \rangle$, $\langle \mathbf{p}_t \rangle$, $\langle
\mathbf{q} \rangle$, $\langle \mathbf{p} \rangle$, and their
moments. The lowest order moments correspond to $k+l+m+n=2$\ (see
below) and are easy to identify as the standard deviations and
covariances in the observables. For example, $\langle \left(
\mathbf{q} - \langle \mathbf{q} \rangle \right)^2 \rangle$\ is
precisely the spread of the wavefunction in the $x_1$\ coordinate.
Using a sharply peaked Gaussian wavefunction in $x_0$\ and $x_1$ it
is straightforward to verify the semiclassical fall-off of the
moments
\[
\left\langle(\mathbf{t}-\langle \mathbf{t} \rangle)^{k}
(\mathbf{p}_t-\langle \mathbf{p}_t \rangle)^{l} (\mathbf{q}-\langle
\mathbf{q} \rangle)^{m} (\mathbf{p} - \langle \mathbf{p}
\rangle)^{n} \right\rangle_{\rm Weyl} \ \ \propto \ \
\hbar^{\frac{1}{2}(k+l+m+n)}\,.
\]

\section{Constraint Functions}

Once the free system has been quantized, Dirac's prescription for
enforcing a constraint is to restrict to the subspace of vector
states that are annihilated by the constraint operator. In other
words, physical wavefunctions are solutions to the equation
\[
\mathbf{C} | \psi \rangle = 0\,.
\]
In many interesting cases, this equation can be solved, but the
solutions have an infinite norm with respect to the kinematical
inner product. Exact techniques, such as group
averaging~\cite{GrpAv_RAQ}, provide a method for \emph{formally}
defining the physical inner product on the space of solutions, with
exact calculations typically being fairly involved.

We observe that, in general, solutions of the quantum constraint
have no kinematical dual and therefore, there is no \emph{a priori}
method for taking expectation values. In fact, there is typically an
infinite number of ways to define a state, in the sense of a linear
functional on $\mathscr{A}$, using such a wave-function. For example
a momentum eigenstate $\exp(i \omega x)$, is not square-integrable
with respect to $x$, however we can define expectation values using
a suitably well-behaved distribution $f(x)$ as
\[
\langle \mathbf{A} \rangle_f := \int {\rm d}x f(x) \mathbf{A} \exp(i
\omega x)\,.
\]
The weak normalization condition would only require that $\int {\rm
d}x f(x) \exp(i \omega x) = 1$, for the given value of $\omega$,
which can be satisfied by a large variety of distributions. The fact
that the state was constructed using an eigenfunction of momentum
manifests itself by momentum acting multiplicatively on such a state
when the corresponding element appears to the right but not on the
left, i.e. $\langle \mathbf{Ap} \rangle_f = \hbar \omega \langle
\mathbf{A} \rangle_f \neq \langle \mathbf{pA} \rangle_f$. By
analogy, we demand that states be annihilated when the constraint
appears on the right, that is
\[
\langle \mathbf{AC} \rangle = 0, \ \ \forall \mathbf{A} \in
\mathscr{A}\,.
\]
Choosing the constraint to annihilate states when appearing on the
left is entirely equivalent. Enforcing this condition leads to an
infinite countable set of constraint functions on the space of
states, which can be imposed systematically by setting $\langle
\mathbf{a}_{\rm 1}^{n_1} \ldots \mathbf{a}_{\rm N}^{n_{\rm N}}
\mathbf{C} \rangle = 0$\ for all values of $n_{\rm 1}, \ldots n_{\rm
N}$. The complete set of these conditions is closed with respect to
the Poisson bracket on $\bar{\mathscr{A}}$, the system is thus
analogous to a classical system with first-class constraints.

The states, such as $\langle \rangle_f$\ above, are not always
positive with respect to the kinematical $\star$-relations, in the
algebraic sense that $\langle \mathbf{A} \mathbf{A}^{\star} \rangle
\geq 0$ will in general not hold for all $\mathbf{A}$. Consequently
the states can assign complex values to expectation values and
moments of $\star$-invariant elements. It is our observation that
kinematically non-positive states are needed in order to impose the
quantum constraint in the above sense. We note, that the states can
still be positive with respect to a subalgebra of $\mathscr{A}$, for
instance the elements that commute with the constraint.

\emph{Free Relativistic Particle.} In terms of wavefunctions in
$x_0$ and $x_1$, $\mathbf{C}|\psi\rangle=0$\ takes the form of the
usual Klein-Gordon equation. Here we impose the quantum constraint
systematically by demanding $\langle \mathbf{t}^k \mathbf{p}_t^l
\mathbf{q}^m \mathbf{p}^n \mathbf{C} \rangle = 0$ for all values of
$k,\ l,\ m,\ n$. We adapt these conditions to semiclassical
expansion by expressing them in terms of expectation values and
moments. Below are examples of two of the quantum constraints
expressed in this way, where short-hand notation was used for the
expectation values $a := \langle \mathbf{a} \rangle$ and the moments
of second order: $(\Delta a)^2 := \langle (\mathbf{a}-a)^2 \rangle$\
and $\Delta(ab) := \langle(\mathbf{a}-a)(\mathbf{b}-b)\rangle_{\rm
Weyl}$
\begin{eqnarray*}
C &=& \langle \mathbf{C} \rangle = p_t^2 - p^2 - m^2 + (\Delta
p_t)^2 - (\Delta p)^2 = 0 \\ C_{t} &=& \langle (\mathbf{t}-\langle
\mathbf{t} \rangle) \mathbf{C} \rangle = 2p_t \Delta(t p_t) + i\hbar
p_t - 2p \Delta(t p) + (3{\rm rd\ order\ moments}) = 0
\end{eqnarray*}
For convenience, we have replaced the constraint function $\langle
\mathbf{tC} \rangle$\ with $\langle (\mathbf{t} - \langle \mathbf{t}
\rangle ) \mathbf{C} \rangle$, which is equivalent when other
constraint functions are taken into account as the two are related
by adding $t \langle \mathbf{C} \rangle$. Other constraint functions
may be expanded similarly.

\section{Gauge Freedom}

As one may recall from analysis of classical constrained systems,
first-class constraints, through Poisson bracket, generate
transformations that preserve the constraint surface and are
regarded as gauge. In our case, infinitesimal gauge transformations
are generated by the action of the constraint element when it
appears on the left, since our constraint condition only demands
that the states are annihilated by the constraint element appearing
to the right. Constraint functions introduced in the previous
section are only meant to be analogous to imposing $\mathbf{C} |
\psi \rangle = 0$. In particular, we have introduced no analogue of
the physical inner product. Consequently, it does not come as a
surprise that we are still left with a number of gauge choices. One
may be puzzled here as the exact Hilbert space constructions, such
as group averaging, typically provide a unique physical inner
product on the space of solutions---where then does this ambiguity
in taking the expectation values come from?

To address this question, we point out that such a physical inner
product is defined only on the space of solutions and does not in
general extend to other wavefunctions in a natural way. As a result,
the expectation values of operators that do not preserve the space
of solutions, are not well-defined. The gauge transformation
generated by a constraint function $\langle \mathbf{AC} \rangle$\ on
the expectation value of some observable $\mathbf{O}$\ is given by
\[
\delta \langle \mathbf{O} \rangle = \{ \langle \mathbf{O} \rangle,
\langle \mathbf{AC} \rangle \} = \frac{1}{i\hbar} \left\langle
\mathbf{A}[\mathbf{O}, \mathbf{C}] \right\rangle + \frac{1}{i \hbar}
\left\langle [\mathbf{O}, \mathbf{A}] \mathbf{C} \right\rangle
\]
It is easy to see that the function generated by an operator that
commutes with the constraint on the solution space, and therefore
preserves it, has a vanishing Poisson bracket with all of the
constraint functions. Therefore, in our formalism, there is also no
ambiguity in taking expectation values of observables that preserve
solutions to the constraint. These are the true physical degrees of
freedom of the system.

To extract invariant quantities one can fix all of the gauge freedom
or solve the differential equations for all the gauge flows. For the
quantities that do not commute with the constraint, one may attempt
to ask how they vary along the gauge orbits \emph{relative to each
other}~\cite{GeomObs1}~\cite{GeomObs2}. This is a physically
meaningful question especially when one is dealing with a so-called
Hamiltonian constraint, which itself should govern the dynamics.
Classically, a Hamiltonian constraint produces a single gauge flow
on the constraint surface. Multiplying the constraint function by
any other function that does not vanish on the constraint surface
results in the same gauge-orbit structure with a different
parametrization. One can take two functions on the constraint
surface and parameterize the gauge flow of one relative to the other
\[
\frac{\rm d}{{\rm d}g} f = \left( \frac{{\rm d}g}{{\rm d} \tau}
\right)^{-1} \frac{\rm d}{{\rm d} \tau} f = \frac{\{ f, C \}}{\{ g,
C \}}\,,
\]
where $\tau$\ is the parameter along the orbits generated by $C$.
This is well-defined so long as the bracket $\{g, C\}$\ does not
vanish. In our quantum analysis, a single Hamiltonian constraint
gives rise to an infinite number of first-class quantum constraint
functions, which generate an infinite number of gauge flows. One
could guess that the flow induced by the expectation value of the
original constraint operator itself is somehow ``preferred''.
However, classically, the constraint function is only fixed up to
multiplication by a non-vanishing function, so that even with
quantization ambiguities set aside, there is no preferred constraint
operator. From the kinematical perspective, any operator with the
same zero-eigenspace as $\mathbf{C}$\ will yield the same set of
solutions, but would in general produce a different flow. For
example if $\mathbf{A}$\ is invertible, kinematically self-adjoint
and commutes with $\mathbf{C}$, the operator product
$\mathbf{C}^{\prime} = \mathbf{AC}$\ is just as good as a constraint
operator but produces a different unitary flow. For our purposes we
assume that all combinations of transformations induced by the
quantum constraint functions are equally viable directions of
evolution. Thus in order to use one of the functions on the
constraint surface as an evolution parameter we first need a way to
reduce the infinite number of gauge directions to a single one.

Below we summarize a method that achieves such a gauge reduction
order-by-order within semiclassical analysis. Let $\mathbf{X}$\ be
an observable chosen to serve as a ``clock''. We would like $\langle
\mathbf{X} \rangle$\ to measure time and the element itself to act
on states simply as a multiplication by $\langle \mathbf{X}
\rangle$. In principle, we once again have the option to demand that
$\mathbf{X}$\ acts as a multiplicative operator when appearing on
the left or on the right. We require that $\langle \mathbf{XA}
\rangle = \langle \mathbf{X} \rangle\langle \mathbf{A} \rangle$ for
all $\mathbf{A}$ as the gauge condition, since it is the only choice
consistent with our constraint functions, namely $\langle
\mathbf{XAC} \rangle = 0 = \langle \mathbf{X} \rangle\langle
\mathbf{AC} \rangle$. The preferred flow is the one that preserves
this gauge. We do not possess a proof or a strong argument that
suggests that these conditions always result in a reduction to one
and only one preferred gauge transformation. In the examples studied
so far, however, we have found that order-by-order in semiclassical
analysis, the method does yield the correct number of gauge fixing
conditions leaving $\langle \mathbf{X} \rangle$\ free to vary along
the single remaining flow.

\emph{Free Relativistic Particle.} For the particle, is natural to
choose the element corresponding to its time coordinate
$\mathbf{t}$---as the evolution variable. In terms of the moments
the gauge conditions are $(\Delta t)^2 = 0$, $\Delta(tp_t) = -
\frac{i \hbar}{2}$, $\Delta(tq) = 0$, $\Delta(tq) = 0$\ as well as
similar conditions for higher order moments.

\section{Semiclassical Analysis}

In order to extract the leading order quantum corrections to the
classical equations of motion we restrict our analysis to the states
that are kinematically semiclassical in the sense defined earlier.
Namely, we assume that M-th order moments have magnitudes of the
order $\hbar^{\frac{1}{2}{\rm M}}$\ and truncate constraint
functions at some chosen power of $\hbar$. For a constraint that is
polynomial in the basic observables, the truncation yields a finite
system of polynomial equations in expectation values and moments.
The set of equations may have more than one solution, each of which
needs to be tested for consistency with the original semiclassical
assumption. The complete infinite set of constraint functions is
first-class, the finite truncated set of constraint functions is
generally only first-class to the relevant order in $\hbar$. Adding
gauge-fixing conditions breaks the first-class nature of the system,
where the residual gauge freedom is given by the linear combinations
of constraint functions that have a vanishing bracket with the gauge
conditions.

The final problem, which can be complicated in general, is ensuring
positivity of the gauge-fixed state with respect to the gauge
invariant algebra elements. In principle it is equivalent to
enforcing reality of the expectation values and quantum uncertainty
relations on moments, however gauge-fixing conditions in general
force us to modify Poisson brackets between variables making them
more difficult to interpret.

\emph{Free Relativistic Particle.} Here we consider a semiclassical
truncation at order $\hbar$---we discard all terms of order
$\hbar^{\frac{3}{2}}$\ and higher, dropping moments above second
order as well as products of second order moments. The truncated
versions of the constraint functions used as an example earlier are
\begin{eqnarray*}
C &=& p_t^2 - p^2 - m^2 + (\Delta p_t)^2 - (\Delta p)^2 = 0
\\ C_t &=& 2p_t \Delta(t p_t) + i\hbar p_t - 2p \Delta(t p)
\end{eqnarray*}
Truncating constraint functions of the relativistic particle one is
left with only five non-trivial constraints and fourteen variables
describing the state up to the second order in moments. The
truncated system of equations has two solutions compatible with the
semiclassical assumption
\begin{eqnarray*}
p_t &=& \pm E \nonumber \\
\Delta(tp_t) &=& \pm \frac{p}{E} \Delta(tp) - \frac{i\hbar}{2}
\nonumber \\ (\Delta p_t)^2 &=& p^2 + m^2 + (\Delta p)^2 - E^2
\nonumber \\ \Delta(p_tq) &=& \pm \frac{p}{E} \left( \Delta(qp) +
\frac{i\hbar}{2} \right) \nonumber \\ \Delta(p_tp) &=& \pm
\frac{p}{E} (\Delta p)^2\,, \\ {\rm where} \quad E &=& \sqrt{p^2 +
m^2} \left( 1 + \frac{m^2 (\Delta p)^2}{ 2( p^2 + m^2)^2} \right)\,.
\end{eqnarray*}
The two constraint surfaces are clearly the extensions of the two
classical solutions $p_t = \pm \sqrt{p^2+m^2}$. We use the above
solutions to completely eliminate the variables generated by
$\mathbf{p}_t$.

The five non-trivial constraints produce only four independent gauge
flows, which ultimately is a consequence of the degeneracy of the
Poisson structure on $\bar{\mathscr{A}}$. In the previous section we
have identified four gauge conditions associated with choosing
$\mathbf{t}$\ as time, it may appear that after gauge-fixing we will
have no flow left to label as evolution along parameter $\langle
\mathbf{t}\rangle$. However a linear combination of these gauge
conditions is already encoded in the constraint $C_t$, since it is
equivalent to the gauge condition $\langle \mathbf{tC} \rangle =
\langle \mathbf{t} \rangle \langle \mathbf{C} \rangle$\ which is
automatically satisfied on the constraint surface. Following through
with the gauge-fixing we find that a single gauge generator $C_{\rm
Ham}$\ remains, its parametrization is fixed by demanding that $\{t,
C_{\rm Ham} \} = 1$. We find that $C_{\rm Ham} = p_t \pm E$, where
the sign depends on the solution chosen. To order $\hbar$\ the
expression for $E$\ agrees with the semiclassical expectation value
of the Klein-Gordon Hamiltonian $\mathbf{H} = \left( \mathbf{p}^2 +
m^2\mathds{1} \right)^{\frac{1}{2}}$.

Finally, we enforce positivity by demanding
\begin{eqnarray*}
&& q, p, (\Delta q)^2, \Delta(qp), (\Delta p)^2 \in \mathds{R} \\
&& (\Delta p)^2, (\Delta q)^2 \geq 0 \\ && (\Delta p)^2 (\Delta q)^2
- \left( \Delta(qp) \right)^2 \geq \frac{1}{4}\hbar^2\,.
\end{eqnarray*}
In this particular example, the gauge conditions do not affect the
Poisson algebra of the the variables generated by $\mathbf{q}$\ and
$\mathbf{p}$\ and the difficulty mentioned earlier is avoided. Given
an initial state we can now evolve its leading order moments for as
long as the semiclassical approximation remains valid. For details
of this example see~\cite{EffConsRel}.

\section{Concluding remarks}

We have developed a technique for deriving leading order
semiclassical corrections for the evolution of quantum systems with
constraints. Within our treatment, a quantum mechanical system is
described by a point on an infinite-dimensional phase-space of
expectation values and moments. Quantum constraint produces an
infinite number of first-class constraint functions. Semiclassical
approximation allows us to truncate the system to a finite
dimensional one at which point the local structure of quantum
constraint surface and gauge orbits can be analyzed using
essentially classical techniques. We have identified a set of gauge
conditions related to the choice of a time variable. Since the
procedure makes use of semiclassical states on kinematical
observables, imposing positivity in general remains complicated. The
scheme is of immediate use in quantizing minisuperspaces such as the
various quantum cosmological models, this venue to be pursued
elsewhere.

\bibliographystyle{plain}

\end{document}